\documentclass[11pt,showpacs,superscriptaddress,amsmath,amssymb,nofootinbib]{revtex4}
\usepackage{graphicx}
\usepackage{dcolumn}
\usepackage{bm}
\usepackage{amssymb}
\usepackage{amsmath}
\usepackage{epsfig}    
\usepackage{color}
\usepackage{slashed}
\usepackage{float}
\newcolumntype{I}{!{\vrule width 1.3pt}}
\begin{document} 
\title{Novel Constraint on Parameter Space of the Georgi-Machacek Model by Current LHC Data}
\preprint{UT-HET-096}
\author{Cheng-Wei Chiang}
\affiliation{Department of Physics and Center for Mathematics and Theoretical Physics, 
National Central University, Chungli, Taiwan 32001, ROC}
\affiliation{Institute of Physics, Academia Sinica, Taipei, Taiwan 11529, ROC}
\affiliation{Physics Division, National Center for Theoretical Sciences, Hsinchu, Taiwan 30013, ROC}
\author{Shinya Kanemura}
\affiliation{Department of Physics, University of Toyama, 3190 Gofuku, Toyama 930-8555, Japan}
\author{Kei Yagyu}
\affiliation{Department of Physics and Center for Mathematics and Theoretical Physics,
National Central University, Chungli, Taiwan 32001, ROC}

\begin{abstract}

The same-sign diboson process $pp\to W^\pm W^\pm jj$ has been measured at the LHC using leptonic decay channels of the $W$ bosons, with production cross sections of two fiducial regions reported to be consistent with the standard model expectations within 1 sigma. 
These results constrain new physics models with a modified $W^+W^+W^-W^-$ vertex.  We consider in particular the Georgi-Machacek model in which the quartic $W$ boson vertex is effectively modified due to mediations of new Higgs bosons in the model.
The relevant gauge-gauge-scalar couplings are all proportional to the vacuum expectation value of the isospin triplets, which can be of $\mathcal{O}(10)$ GeV because of custodial vacuum alignment.  
Using the current 8-TeV data at the LHC, we exclude parameter space on the plane of the triplet vacuum expectation value and the new Higgs boson mass.  
The expected discovery reach at the 14-TeV LHC is also studied. 

\pacs{12.60.Fr, 14.80.Fd}

\end{abstract}
\maketitle
\newpage

\section{Introduction \label{sec:intro}}
%
Since its discovery in 2012 summer, the 126-GeV Higgs boson has attracted a lot of interest in studying its detailed properties.  
Most of such studies are trying to address the question whether it is exactly the one predicted by the Standard Model (SM) 
or, equivalently, whether it is one of scalar bosons in models with an extension of the Higgs sector. 
An extended Higgs sector generally leads to a modification of cross sections of various SM processes 
due to an effect of additional Higgs bosons. 
It is therefore an important program to study how we can extract the structure of the Higgs sector from modified cross sections.

Recently, the ATLAS Collaboration presented the first measurement of the $pp\to W^\pm W^\pm jj$ process with leptonic decays of the $W$ bosons based on their $20.3$ fb$^{-1}$ collision data at $\sqrt{s} = 8$ TeV~\cite{ATLAS_SSW}.  They determined the production cross sections in both inclusive and vector boson scattering (VBS) fiducial regions, with the latter being a subset of the former and more sensitive to electroweak processes.  
Such data provide constraints on the size of effective $W^+W^+W^-W^-$ vertex that can be significantly
modified by mediations of doubly-charged Higgs bosons $H^{\pm\pm}$; i.e., $W^\pm W^\pm\to H^{\pm\pm}\to W^\pm W^\pm$ through the $W^\pm W^\pm H^{\mp\mp}$ couplings. 

The magnitude of $W^\pm W^\pm H^{\mp\mp}$ couplings is proportional to the vacuum expectation value (VEV) of an additional Higgs multiplet with a representation larger than the isospin doublet.  Usually, such a VEV is constrained to be smaller than a few GeV by the electroweak $\rho$ parameter, as is the case in, {\it e.g.}, the minimal Higgs triplet model~\cite{HTM}.  
In this case, the modification in the quartic $W$ boson coupling is sufficiently small to be neglected. 
However, this constraint is circumvented in the Georgi-Machacek (GM) model~\cite{GM} in which a pair of complex and real triplet fields with vacuum alignment are introduced, thereby preserving the custodial symmetry and permitting an {\cal O}(10) GeV triplet VEV\footnote{If this alignment of 
triplet VEV's is relaxed, the upper bound on the triplet VEV's can be similar to that in the minimal triplet Higgs model. 
In such a case, the $W^\pm W^\pm\to H^{\pm\pm}\to W^\pm W^\pm$ process is negligible, 
and the pair production $pp\to H^{++}H^{--}$ and the associated production with singly-charged Higgs bosons $pp\to H^{\pm\pm}H^\mp$ can be used for the $H^{\pm\pm}$ production at the LHC. 
From the same-sign dilepton data at the LHC with the collision energy to be 7 TeV,  
the lower bound for the mass of $H^{\pm\pm}$ has been derived to be about 400 GeV~\cite{dilepton} and about 60 GeV~\cite{diboson} when 
$H^{\pm\pm}$ mainly decay into the same-sign leptons and dibosons, respectively. }.  
Even though a Higgs sector with an isospin septet field also maintains $\rho = 1$ at tree level~\cite{septet}, 
it has an accidental U(1) symmetry that could lead to an unwanted Nambu-Goldstone (NG) boson once it is spontaneously broken.  
We therefore consider the GM model as the minimal well-defined model.

In addition to large $W^\pm W^\pm H^{\mp\mp}$ vertices, such a large triplet VEV also gives rise to sizeable $ZW^\pm H^\mp $ vertices, where $H^\pm$ are physical singly-charged Higgs bosons.  
The $ZW^\pm H^\mp $ vertices are known to be small in multi-doublet models, because they appear at loop levels~\cite{multi1,multi2,multi3}.   
The size of the vertices is also suppressed in the minimal Higgs triplet model due to the small triplet VEV, similar to the case of $W^\pm W^\pm H^{\mp\mp}$ vertices. 
Thus, the measurement of $ZW^\pm H^\mp$ vertices is another probe of the GM model. 
The feasibility studies of measuring the $ZW^\pm H^\mp$ vertices have been done 
in Refs.~\cite{HWZ_LHC1,HWZ_LHC2,HWZ_LHC3} for the LHC, and in Refs.~\cite{HWZ_ILC1,HWZ_ILC2} for future linear colliders.

Furthermore, the large triplet VEV deviates the coupling constants for the SM-like Higgs boson $h$ with the weak gauge bosons $hVV$ and the fermions $hf\bar{f}$ from their SM values. 
Unlike the Higgs sector composed of only singlets and doublets, 
the $hVV$ couplings can be larger than the SM predictions as a result of the larger isospin representation, serving as a distinctive feature of the GM model.

In this work, we compare the cross section of the VBS region driven by the data with its prediction in the GM model. 
We then show that a significant portion of the parameter space has been ruled out by the ATLAS data. 
We also estimate the discovery reach for the new Higgs bosons using the same process at the 14-TeV LHC.  Based on the above constraint, one can find an upper bound on the triplet VEV for each given mass of the new Higgs bosons.  
This upper bound in turn determines possible ranges of the $hVV$ and $hf\bar{f}$ coupling constants and thus 
the signal strengths for the $gg\to h\to VV$ modes in a very predictive way. 

This paper is organized as follows.  
The GM model is briefly reviewed in Section~\ref{sec:GM}.  
Section~\ref{sec:LHC} shows our primary results based on the latest ATLAS data.  
Our findings are summarized in Section~\ref{sec:summary}.

\section{Georgi-Machacek Model \label{sec:GM}}
%
The Higgs sector in the GM model is composed of an isospin doublet field, $\phi$, with hypercharge $Y=1/2$, 
a complex triplet field, $\chi$, with $Y=1$, and a real triplet field, $\xi$, with $Y=0$, 
where the electric charge $Q$ is given by $Q=T_3+Y$ with $T_3$ being the third component of the isospin.
These fields can be expressed in the $SU(2)_L\times SU(2)_R$ covariant form as:
\begin{align}
\hspace{-5mm}\Phi=
\begin{pmatrix}
\phi^{0*} & \phi^+ \\
-(\phi^+)^* & \phi^0
\end{pmatrix},~
\Delta=
\begin{pmatrix}
\chi^{0*} & \xi^+ & \chi^{++} \\
-(\chi^+)^* & \xi^0 & \chi^{+} \\
(\chi^{++})^* & -(\xi^+)^* & \chi^{0} 
\end{pmatrix}. 
\label{eq:Higgs_matrices}
\end{align}
The neutral components are parameterized as 
\begin{align}
\phi^0&=\frac{1}{\sqrt{2}}(v_\phi+\phi_r+i\phi_i), \notag\\
\chi^0&=v_\chi+\frac{1}{\sqrt{2}}(\chi_r+i\chi_i),\quad \xi^0=v_\xi+\xi_r, \label{eq:neutral}
\end{align}
where $v_\phi$, $v_\chi$ and $v_\xi$ are the VEV's of $\phi$, $\chi$ and $\xi$, respectively.  
Under the $SU(2)_L\times SU(2)_R\times U(1)_Y$ symmetry, the most general Higgs potential contains four dimensionful and five dimensionless parameters, and its explicit form can be found, for example, in Ref.~\cite{Chiang_Yagyu_GM}.

When the two triplet fields develop aligned VEV's, $v_\chi=v_\xi \equiv v_\Delta$, 
the $SU(2)_L\times SU(2)_R$ symmetry reduces to the custodial $SU(2)_V$ symmetry. 
In that case, the $W$ boson mass $m_W$ and the $Z$ boson mass $m_Z$ 
have the same form as in the SM; i.e., 
$m_W^2 = g^2v^2/4$ and $m_Z^2=g^2v^2/(4\cos^2\theta_W)$, 
where $v^2$ is related to the Fermi constant $G_F$ by $v^2\equiv v_\phi^2+8v_\Delta^2=1/(\sqrt{2}G_F)$, and $\theta_W$ is the weak mixing angle. 
Therefore, the electroweak $\rho$ parameter defined by $\rho_{\text{tree}} \equiv m_W^2/(m_Z^2\cos^2\theta_W)$ is unity at tree level. 

The component scalar fields can be classified into irreducible representations of $SU(2)_V$ multiplets. 
The scalar fields from the doublet $\Phi$ are decomposed as ${\bf 2}\otimes {\bf 2}={\bf 3}\oplus{\bf 1}$, and those from the triplet $\Delta$ as ${\bf 3}\otimes {\bf 3}={\bf 5}\oplus{\bf 3}\oplus{\bf 1}$. 
As a result, we have one 5-plet, two 3-plet and two singlet representations, where the latter two can mix with each other with mixing angles $\theta_H$ and $\alpha$. 
The mixing angle $\theta_H$ is defined by $\sin\theta_H=2\sqrt{2}v_\Delta/v$, while $\alpha$ is determined by the quartic coupling constants in the Higgs potential. 
The 5-plet states are physical and denoted by ($H_5^{\pm\pm},H_5^\pm,H_5^0$). 
On the other hand, one linear combination of the two 3-plets corresponds to the NG bosons ($G^\pm,G^0$) that become the longitudinal components of the $W$ and $Z$ bosons, respectively. 
The other linear combination gives the physical 3-plet Higgs bosons ($H_3^\pm,H_3^0$). 
Furthermore, after diagonalizing the mass matrix for the two $SU(2)_V$ singlet fields, 
we obtain two CP-even neutral Higgs bosons. 
One of them should be identified as the 126-GeV Higgs boson at the LHC.
Due to the custodial symmetry, Higgs bosons belonging to the same $SU(2)_V$ multiplet are degenerate in mass. 

Seven of the nine parameters in the Higgs potential can be converted into the following physical parameters: 
four mass parameters for the 5-plet, 3-plet and two singlet Higgs bosons, one mixing angle $\alpha$, and 
two VEV's $v$ and $v_\Delta$. 
The remained two parameters, which are given as the coefficients of the trilinear scalar vertices, 
enter the mass formulas of the Higgs bosons. 
One can tune the two parameters to realize the decoupling limit, 
where the 5-plet, 3-plet and one of the singlet Higgs bosons are much heavier than the electroweak scale.
These parameters can be constrained by such theoretical arguments as perturbative unitarity~\cite{Aoki_Kanemura} 
and vacuum stability~\cite{Logan}.

In the following, we summarize a few characteristic features of the GM model. 
First, the triplet VEV $v_\Delta$ can be $\mathcal{O}$(10) GeV, with an upper limit of about 80 GeV as required by the perturbativity of the top Yukawa 
coupling.  
Such a large triplet VEV gives rise to distinctive phenomenology in collider experiments~\cite{Chiang_Yagyu_GM,Nomura,An-Li,Englert}.

Second, the $hVV$ couplings ($V=W,Z$) for the SM-like Higgs boson $h$ can be larger than their SM values. 
The ratios of the $hVV$ and $hf\bar{f}$ couplings to their respective SM values are given by 
\begin{align}
c_{hVV}^{}=c_H c_\alpha +\sqrt{\frac{8}{3}}s_Hs_\alpha , ~
c_{hff}^{}=\frac{c_\alpha}{c_H}, 
\label{eq:cratio}
\end{align}
where $c_H$, $s_H$, $c_\alpha$ and $s_\alpha$ are respectively $\cos\theta_H$, $\sin\theta_H$, $\cos\alpha$ and $\sin\alpha$. 
It is the group factor $\sqrt{8/3}$ in $c_{hVV}$ that makes it possible for the $hVV$ couplings to be larger than their SM values.  
This feature is not shared by models extended with only multi-doublets and/or singlets.  Therefore, $c_{hVV} > 1$ is one crucial signature to identify the GM model, particularly when $v_\Delta$ is sizeable. 

Finally, the 3-plet Higgs bosons are ``fermion-philic"; {\it i.e.}, they couple to fermions due to their components from $\Phi$ through mixing, but they do not couple to the weak gauge bosons. 
In fact, properties of the 3-plet Higgs bosons are quite similar to those of the CP-odd and singly-charged Higgs bosons in the Type-I two Higgs doublet model. 
Its phenomenology has been studied a lot in the literature~\cite{Type-I}. 
On the contrary, the 5-plet Higgs bosons are ``gauge-philic''; namely, they have interactions with the weak gauge bosons, but they do not couple to SM fermions because they involve purely components in the $\Delta$ field.
Thus, they decay dominantly into gauge boson pairs~\footnote{When the mass of the 3-plet Higgs bosons is smaller than that of the 5-plet Higgs bosons, 
the latter can decay into the former and a weak gauge boson.}. 
Magnitudes of the 5-plet Higgs couplings with the weak bosons are all proportional to $v_\Delta$:
\begin{align}
&(H_5^{\pm\pm}W^\mp W^\mp): \frac{g}{\sqrt{2}}m_Ws_H,~
(H_5^{\pm}W^\mp Z): -gm_Zs_H,\notag\\
&(H_5^{0}W^+W^-): -\frac{gm_Ws_H}{\sqrt{3}},~
(H_5^{0}ZZ): \frac{2 gm_Zs_H}{\sqrt{3}c_W}. 
\end{align}
When $v_\Delta$ is large, cross sections of the VBS processes can be significantly modified due to 
the mediation of the 5-plet Higgs bosons through the above vertices. 
We therefore expect the cross section measured by ATLAS to impose a strong constraint on the model.

\section{Current LHC Constraint \label{sec:LHC}}

For the same-sign diboson process of $pp\to jjW^\pm W^\pm$ with $W^\pm \to \ell^\pm \nu$ ($\ell=e,\mu$) measured by the ATLAS~\cite{ATLAS_SSW}, the 5-plet Higgs bosons $H_5^{\pm\pm}$ and $H_5^0$ can contribute through $s$-channel and $t$-channel, respectively. 
For definiteness, we fix the $hVV$ couplings at their SM values.  We have checked that the VBS cross section varies by less than 1\% when the $hVV$ couplings are changed by $30\%$ and, therefore, the bound on $v_\Delta$ is 
essentially unchanged within the range. 
Under this assumption, the modification of the VBS cross section of $pp\to jjW^\pm W^\pm$ process depends only on $v_\Delta$ and $m_{H_5}$, the mass of $H_5^{\pm\pm}$ and $H_5^0$. 

In Ref.~\cite{ATLAS_SSW}, the signal events are classified as the inclusive region and the VBS region. 
In both of the cases, the following basic kinematic cuts are imposed:
\begin{align}
&p_T^\ell > 20~\text{GeV},~p_T^j > 30~\text{GeV},~E_T\hspace{-4.0mm}/\hspace{2mm}>40~\text{GeV},\notag\\
&|\eta^\ell|<2.5,~|\eta^j|<4.5,\notag\\
&\Delta R_{\ell\ell}>0.3,~\Delta R_{jj} > 0.4,~\Delta R_{\ell j} > 0.3,\notag\\
&M_{jj} > 500~\text{GeV},~M_{\ell\ell} > 20~\text{GeV}, \label{basic}
\end{align} 
where $p_T^X$, and $\eta^X$ and $M_{XX}$ are the transverse mass and pseudorapidity for $X$, respectively. 
The distance between $X$ and $Y$ is denoted by $\Delta R_{XY}$, and $E_T\hspace{-4mm}/\hspace{2mm}$ is the missing transverse energy. 
The signal events for the inclusive region are obtained by only taking the above cuts. 
For the VBS region, one further imposes the following cut:
\begin{align}
|\Delta y_{jj}| > 2.4, 
\label{eq:rapidity}
\end{align}
where $\Delta y_{jj}$ is the rapidity difference between the dijets. 
We note that the cross section of the inclusive region includes contributions from both electroweak and strong processes, while 
that of the VBS region mainly the electroweak processes due to the cut in Eq.~(\ref{eq:rapidity}). 

From the measured $pp\to jj\ell^\pm \ell^\pm E_T\hspace{-4mm}/\hspace{2mm}$ events and Monte Carlo background simulations, the fiducial cross sections for the inclusive and VBS regions are respectively derived to be 
2.1$\pm$0.5(stat)$\pm$0.3(sys) fb and 1.3$\pm$0.4(stat)$\pm$0.2(sys) fb~\cite{ATLAS_SSW}.
The corresponding SM cross sections quoted in Ref.~\cite{ATLAS_SSW} are 1.52$\pm$ 0.11 fb and 0.95$\pm$0.06 fb. 
Therefore, the SM predictions are consistent with the measured fiducial cross sections within 1$\sigma$. 

In the following numerical analysis, we use {\tt MadGraph5}~\cite{MG5} for simulations and {\tt CTEQ6L} for the parton distribution functions. 
Before comparing the cross sections in the GM model with the fiducial values, we first calibrate
the SM cross sections. 
Our SM simulations give the inclusive cross section as 1.66 fb and the VBS cross section as 1.06 fb.  
We will thus multiply the factors 0.92~(=1.52 fb/1.66 fb) and 0.90~(=0.95 fb/1.06 fb) to the cross sections 
simulated in our analysis in the inclusive and VBS regions, respectively.  
%
We confirm that the VBS region has a better sensitivity than the inclusive region.  
For example, using the analysis based on the VBS (inclusive) region, we obtain in the case of $m_{H_5}=200$ GeV the upper limit of 27 GeV (32 GeV) at the 68\% CL and 33 GeV (40 GeV) at the 95\% CL for $v_\Delta$. 
Therefore, we concentrate on the VBS cross section in the following analysis. 

\begin{figure*}[t!]
\begin{center}
\includegraphics[width=75mm]{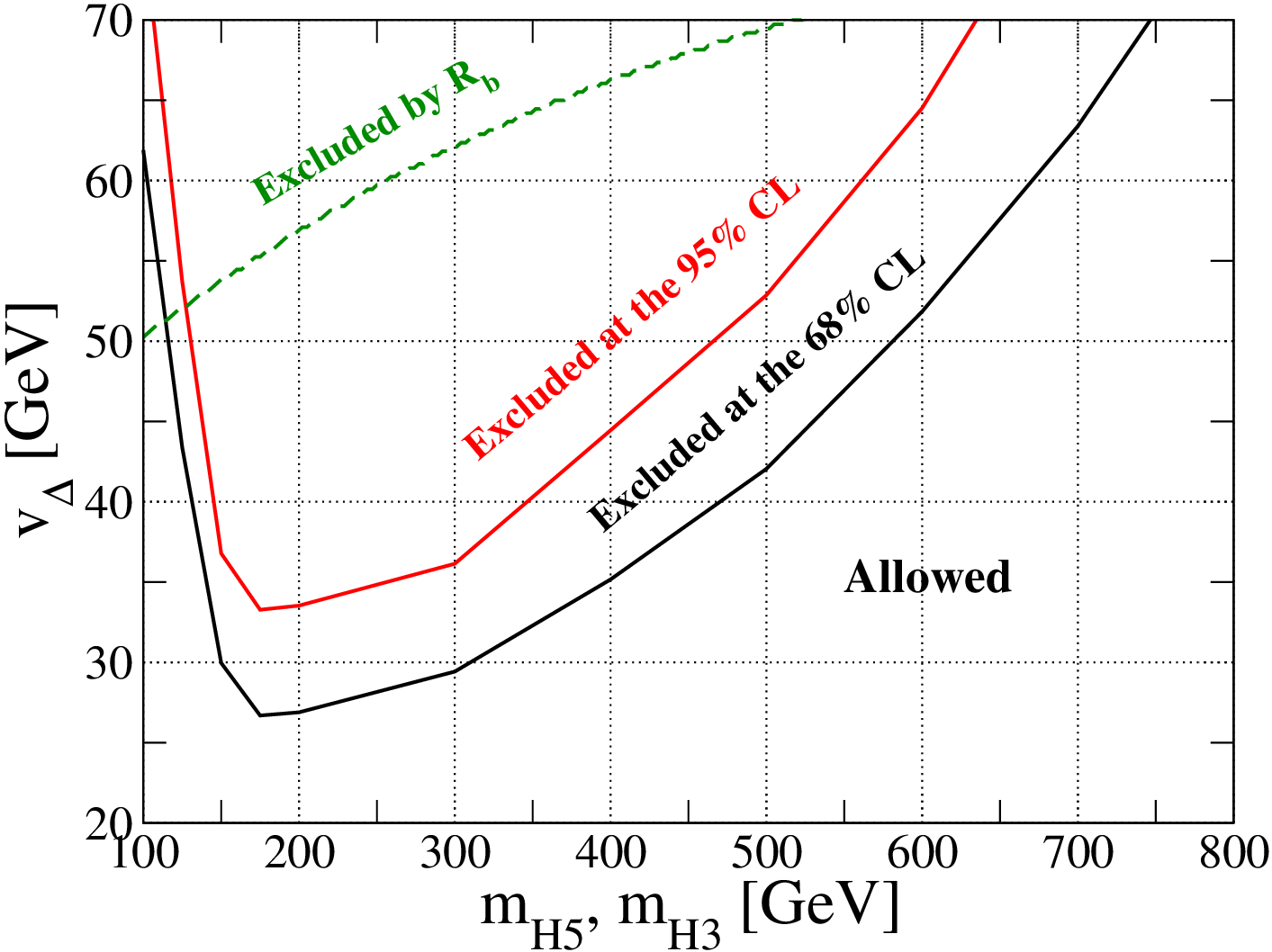}\hspace{7mm}
\includegraphics[width=75mm]{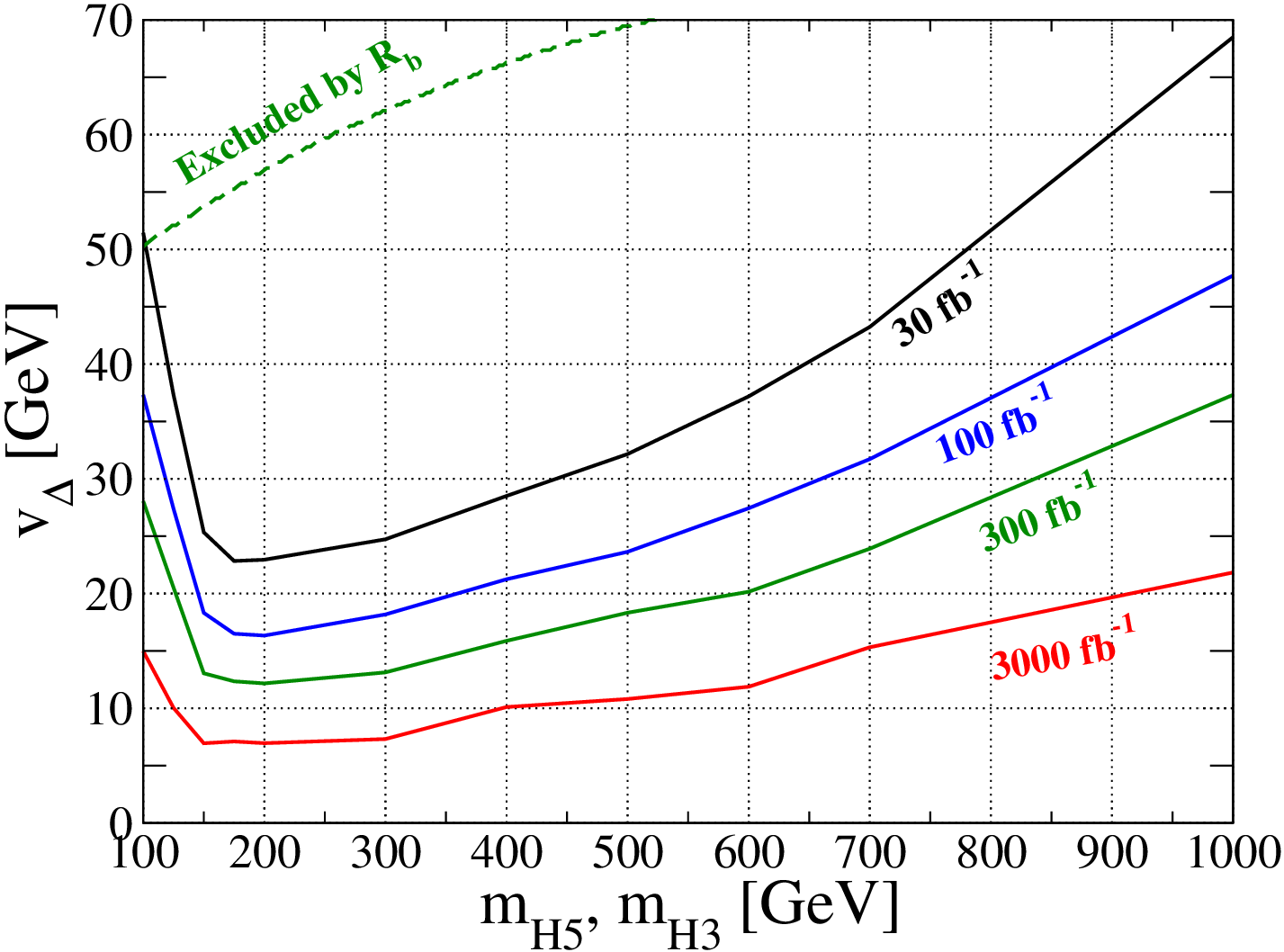}
\end{center}
\caption{
(Left) Excluded regions on the $m_{H_5}$-$v_\Delta$ plane by the 8-TeV LHC data at 68\% and 95\% CL. 
(Right) Contours of required luminosity for a 5-sigma discovery at the 14-TeV LHC on the $m_{H_5}$-$v_\Delta$ plane. 
As a comparison, regions above the dashed curves in both plots are excluded by the $R_b$ data at 95\% CL, where the horizontal axis should be read as the mass of 3-plet Higgs bosons, $m_{H_3}$.
}
\label{Fig:VBA}
\end{figure*}

The left plot in Fig.~\ref{Fig:VBA} shows the excluded parameter region on the $m_{H_5}$- $v_\Delta$ plane according to the current 20.3 fb$^{-1}$ data of 8-TeV LHC.  The region above the black (red) curve is excluded at the 68\% (95\%) CL.  The most severe upper bound on $v_\Delta$ is about 30 GeV at the 95\% CL in the case of $m_{H_5}=200$ GeV.  
When a larger value of $m_{H_5}$ is taken, the bound on $v_\Delta$ becomes more relaxed due to smaller production cross sections. 
When $m_{H_5}$ is taken to be smaller than about 200 GeV, a milder bound on $v_\Delta$ is also obtained, as more events from the 5-plet Higgs bosons are rejected by the kinematic cuts in Eq.~(\ref{basic}). 
As a comparison, we also show by the green dashed curve in this plot the constraint from the measurement of $R_b^{\text{exp}}=0.21629\pm 0.00066$~\cite{PDG}, which depends on the mass of 3-plet Higgs bosons $m_{H_3}$ and $v_\Delta$~\cite{Haber,Chiang_Yagyu_GM}. The region above the dashed curve is excluded at 95\% CL.

By applying the same analysis for the VBS region to the case of 14-TeV collisions, 
one can calculate expected cross section deviations from the SM predictions for different luminosities. 
In the right plot of Fig.~\ref{Fig:VBA}, we show the expected 5-sigma reach for excess in the $pp\to jj W^\pm W^\pm$ process at the 14-TeV LHC on the $m_{H_5}$- $v_\Delta$ plane. 
The integrated luminosity is assumed to be 30, 100 and 300 fb$^{-1}$ for the three curves. 
Similar to the analysis of 8-TeV data, the discovery reach becomes the largest at around $m_{H_5}=200$ GeV, where 
a 5-sigma discrepancy is expected in the cases of $v_\Delta \gtrsim 24$, 17 and 12 GeV for the luminosity of 30, 100, 300, and 3000 fb$^{-1}$, respectively. 
The constraint from $R_b$ is also displayed in this plot.

\begin{figure}[t]
\hspace{-1cm}
\includegraphics[width=90mm]{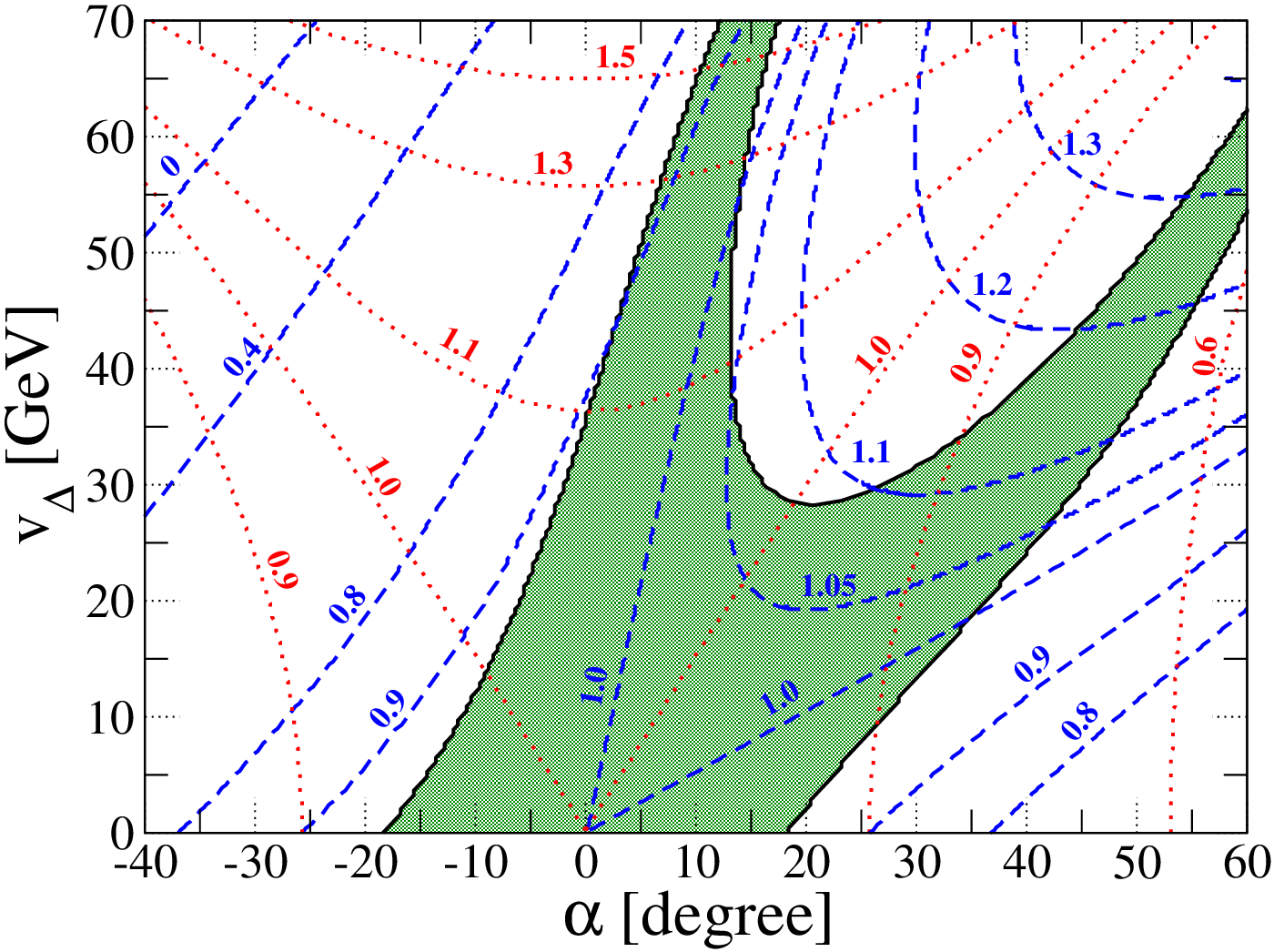}
\caption{
The green region on the $\alpha$-$v_\Delta$ plane marks the region where $\mu_{VV}$, defined in Eq.~(\ref{mu_VV}), are within $1.0\pm 0.1$.  The contours for the $hVV$ coupling ratio, $c_{hVV}$, and those for the $hff$ coupling ratio, $c_{hff}$, are also superimposed as blue dashed and red dotted curves, respectively. 
}
\label{Fig:mu_VV}
\end{figure}

Fig.~\ref{Fig:mu_VV} shows the contours of signal strengths for the $h\to VV$ channel defined by  
\begin{align}
\mu_{VV}^{}\equiv 
\frac{\sigma(gg\to h)\times \mathcal{B}(h\to VV)}{\sigma(gg\to h)_\text{SM}\times \mathcal{B}(h\to VV)_\text{SM}}, \label{mu_VV}
\end{align}
from the gluon fusion production mechanism, where $\sigma(gg\to h)$ and $\mathcal{B}(h\to VV)$ are respectively the gluon fusion cross section and the branching ratio of the $h\to VV$ decay. 
It is seen that even in the case with $\mu_{VV}^{}\simeq 1$ as favored by the current LHC data, 
there are regions with $c_{hVV}^{}>1$.  
This is because the enhancement in $\mathcal{B}(h\to VV)$ is compensated by the suppression in the gluon fusion cross section due to $c_{hff}^{}<1$. 
This signifies the importance of studying the VBS process in addition to the gluon fusion process in order to reliably fix the exact values of $hVV$ and $hff$ couplings. Signatures of enhanced VBS processes in the GM model at the LHC had been studied in Ref.~\cite{An-Li}.

Finally, we comment on precision measurements of the $hVV$ couplings at future collider experiments such as the High-Luminosity LHC (HL-LHC) and the International Linear Collider (ILC).
At the ILC, the $hVV$ couplings are supposed to be measured at about 1 \% accuracy~\cite{ILC_TDR,ILC_White} with the 500-GeV collisions and an integrated luminosity of 500 fb$^{-1}$.
By combining the independent measurements of $\mu_{VV}^{}$ at the HL-LHC and the $hVV$ couplings at the ILC, 
we will be able to determine parameters of the GM model, such as $\alpha$ and masses of extra Higgs bosons~\footnote{
When $c_{hVV}\neq 1$ is found in future experiments, one can obtain upper limits on the masses of extra Higgs bosons from the unitarity and vacuum stability bounds~\cite{An-Li}. } even if we cannot directly discover them at the early stage.

\section{Conclusions \label{sec:summary}}

We have studied the constraint on the GM model based on the first measurement of the same-sign diboson process $pp\to W^\pm W^\pm jj$ by the ATLAS Collaboration.  In the GM model, the doubly-charged and neutral Higgs bosons in the $SU(2)_V$ 5-plet contribute to the above process.  
Because the gauge-gauge-scalar vertices for the 5-plet Higgs bosons are all proportional to the triplet VEV $v_\Delta$, we have established the excluded parameter space on the $m_{H_5}$-$v_\Delta$ plane. 
For example, we have found that the upper bound on $v_\Delta$ is about 30, 37 and 55 GeV for the case of $m_{H_5}=200$, 400 and 600 GeV at the 95\% CL, respectively. 
We have also analyzed the expected 5-sigma reach for the same process at the 14-TeV LHC. 
As an example, assuming the integrated luminosity of 300 fb$^{-1}$, we expect a 5-sigma deviation for $v_\Delta=$ 12, 19 and 38 GeV when $m_{H_5}=200$, 500 and 1000 GeV, respectively.

\section*{Acknowledgments}
S.K. thanks the hospitality of NCTS, Taiwan during his visit where part of this work was done.
%
This work was supported in part by the Ministry of Science and Technology of R.~O.~C. 
under Grant Nos.~NSC-100-2628-M-008-003-MY4, NSC-101-2811-M-008-014, the National Science Foundation under Grant No.~NSF~PHY11-25915, and Grant-in-Aid for Scientific Research from JSPS and MEXT, Nos. 22244031,  24340046 and 23104006.

\end{document}